# Polycrystalline graphene and other two-dimensional materials

Oleg V. Yazyev[1,*] and Yong P. Chen[2,#]

[1] Institute of Theoretical Physics, Ecole Polytechnique Fédérale de Lausanne (EPFL), CH-1015 Lausanne, Switzerland

[2] Department of Physics and School of Electrical and Computer Engineering and Birck Nanotechnology Center, Purdue University, West Lafayette, Indiana 47907, USA

[*] To whom correspondence and requests for materials should be addressed. e-mail: oleg.yazyev@epfl.ch

[#] e-mail: yongchen@purdue.edu

**Graphene, a single atomic layer of graphitic carbon, has attracted intense attention due to its extraordinary properties that make it a suitable material for a wide range of technological applications. Large-area graphene films, which are necessary for industrial applications, are typically polycrystalline, that is, composed of single-crystalline grains of varying orientation joined by grain boundaries. Here, we present a review of the large body of research reported in the past few years on polycrystalline graphene. We discuss its growth and**

**formation, the microscopic structure of grain boundaries and their relations to other types of topological defects such as dislocations. The review further covers electronic transport, optical and mechanical properties pertaining to the characterizations of grain boundaries, and applications of polycrystalline graphene. We also discuss research, still in its infancy, performed on other 2D materials such as transition metal dichalcogenides, and offer perspectives for future directions of research.**The field of research in two-dimensional (2D) materials has been enjoying spectacular growth during the past decade. This activity was triggered by pioneering works on graphene[1,2,3], a 2D semimetallic allotrope of carbon that turned out to be an exceptionally fertile ground for advancing frontiers of condensed matter physics[4,5,6,7]. However, the center of interests has rapidly shifted from fundamental science to potential technological applications of this 2D material[8,9,10]. Furthermore, other atomically thin monolayer systems, which possess some very valuable properties for many applications, have soon joined the field thus extending the palette of available 2D materials. Examples are insulating monolayer hexagonal boron nitride ($h$-BN)[11] and semiconducting transition metal dichalcogenides (TMDCs) $MX_2$ (M = Mo, W; X = S, Se) characterized by electronic band gaps between 1.1 eV and 1.9 eV[12,13]. The diversity of 2D materials further opens the possibility for such atomically thin crystals to be combined in complex heterostructures by stacking them on top of each other, thus giving rise to a whole new paradigm of nanoscale engineering[7,14,15,16].

Technological applications require scalable techniques that would produce large-area sheets beyond the micrometer size samples of graphene used in earlier research, such as those single-crystalline graphene flakes readily obtained by mechanical exfoliation of graphite[1]. Statistical physics arguments, however, suggest that crystalline order in 2D is highly susceptible to various types of fluctuation and disorder[17], which would hinder producing high-quality single-crystalline graphene sheets of arbitrarily large size. Practically, typical films of graphene of wafer scale or larger size as produced by, for example chemical vapor deposition (CVD), are polycrystalline[18,19,20], i.e. composed of single-crystalline domains of varying lattice orientation. In polycrystalline materials, such rotational disorder necessarily leads to the presence of grain boundaries (GBs) – interfaces between single-crystalline domains[21,22]. GBs represent a class of topological defects – imperfections described by a structural topological invariant that does not change upon local modifications of the lattice[23]. Of course, such topological defects, intrinsic to polycrystalline materials, inevitably affect all properties of the material under study.

This Review discusses recent experimental advances of the emerging field of polycrystalline 2D materials complemented with necessary theoretical concepts. We first cover recent progress in observing the micrometer scale morphology in polycrystalline graphene as well as the atomic structure of GBs. The structure of the latter is explained in terms of hierarchical classification of topological defects in crystalline lattices. Special attention is devoted to peculiar behavior of topological defects in 2D graphene as opposed to those in bulk 3D crystals. The article then covers important aspects of graphene growth by CVD for the formation of polycrystalline graphene. We then consider electronic transport, optical, mechanical and thermal properties and related characterization techniques of polycrystalline graphene. The final section reviews several polycrystalline 2D materials other than graphene – monolayer hexagonal boron nitride (*h*-BN), TMDCs such as $MoS_2$, and 2D silica. The Review is concluded with an outlook of future directions of research in this field.

**Structure of polycrystalline graphene**

**Experimental evidence.** Historically, research on polycrystalline graphene was preceded by investigations of topological defects in bulk graphite. First transmission electron microscopy (TEM) studies of dislocations in graphite were reported in early sixties[24,25]. In 1966, Roscoe and Thomas proposed an atomistic model of tilt GBs in graphite, suggesting that the cores of edge dislocations are composed of pentagons-heptagons pairs[26], which is consistent with the structure of topological defects in polycrystalline monolayer graphene discussed below. Later, the interest in GB defects in graphene was renewed with the advent of scanning tunneling microscopy (STM) for investigating surfaces[27,28,29,30]. Scanning tunneling spectroscopy (STS) allowed investigating in detail the local electronic properties of these defects in graphite[31]. The scanning probe techniques have also been used recently to explore the possible role of GBs in the intrinsic ferromagnetism of graphite[32].

In 3D bulk solids, the structure of dislocations and GBs is generally difficult to access and image using current microscopy techniques as these defects are mostly buried deep inside. In contrast, 2D materials such as graphene provide an exceptional experimental system where such structural irregularities are exposed and can be studied in greater details by microscopy with resolution down to atomic levels, and even including temporal evolution. For polycrystalline graphene, transmission electron microscopy (TEM) has become one of the most powerful and widely used tools to map out both the polycrystalline morphology on a large scale (i.e. above the size of single-crystalline grains), and the structural details of individual topological defects down to atomic scale[33]. While early TEM observation of a dislocation in graphene has been reported by Hashimoto *et al.* in 2004[34], the first systematic investigations of GBs in polycrystalline graphene were published only in 2011[18,19,20]. These experiments were performed on graphene grown by CVD on Cu substrate[35]. Huang *et al.* employed diffraction-filtered dark field (DF) TEM for large-area mapping of the location, size, orientation and shape of several hundred grains and grain boundaries[18]. In their study, individual crystalline orientations are isolated using an aperture to select the appropriate diffraction spot. The resulting images revealed an intricate patchwork of grains connected by tilt GBs (Figs. 1a–c). The grains in graphene samples produced by Huang *et al.*[18] are predominantly of submicrometer size (Fig. 1d) while GB misorientation angles show a complex multimodal distribution (Fig. 1e). The distribution of grain sizes and misorientation angles, however, depends strongly on the synthetic protocol used for producing graphene. For instance, An *et al.* reported a different distribution of misorientation angle, which is mostly confined between 10 and 30 degrees[20].

Using high-resolution TEM, the details of atomic-scale structure of GB defects have been determined[18,19]. Many GBs exhibited atomically sharp interface regions formed by chains of alternating pentagons and heptagons embedded in the hexagonal lattice of graphene (Fig. 1c), in full agreement with previous theoretical predictions[36,37]. This structure can be understood in light of the Read-Shockley model[38], which views tilt GBs as arrays of edge dislocations. Dislocations in graphene are represented by pairs of pentagons and heptagons (disclinations) – the elementary structural topological defects in graphene. Hierarchical relations between the above-mentioned classes of structural topological defects[23,36,37,39,40] and the definitions of their topological invariants are explained in Box 1. Importantly, this construction based on pentagonal and heptagonal units conserves the coordination environment of all carbon atoms, thus automatically resulting in energetically favorable structures. In contrast, HR-TEM images of An *et al.* show the presence of undercoordinated atoms ("twinlike" structures) in the GB regions[20], likely to be stabilized by adsorbates found in almost all of the boundary areas. The models of GBs containing undercoordinated carbon atoms, either with dangling bonds or forming complexes with extrinsic adsorbates, have been investigated theoretically[41,42]. Besides GBs involving interatomic bonds across the interface region, several studies have also reported weakly connected GBs formed by "overlapping" individual grains, i.e. with one domain grow over the top of a neighboring domain[43,44].

Several examples of topologically trivial defects (that is, characterized by zero values of the relevant structural topological invariants) derived from GBs in graphene deserve special attention. Lahiri *et al.* reported an observation of highly regular line defects in graphene grown on Ni(111) substrate[45]. Such a one-dimensional defect formed by alternating octagons and pentagon pairs aligned along zigzag direction (Fig. 1f) can be viewed as a degenerate GB as it has zero misorientation angle. Because of its topologically trivial structure, this defect can be engineered in a controlled way as demonstrated by Chen *et al.*[46] Another work observed a different line defect in graphene oriented along the armchair direction[47]. GB loops are formally equivalent to point defects in crystal lattices. A striking example is the highly symmetric flower-shaped defect found in graphene produced using different methods (Fig. 1g)[48,49]. Less symmetric small grain boundary loops have also been observed in TEM studies[50,51,52].

Finally, a different type of topological defects is possible in multilayer systems such as bilayer graphene. Several groups have reported observations of boundaries between domains with structurally equivalent AB and AC stacking orders in bilayer graphene[53,54,55,56,57]. These stacking domain boundaries observed by means of DF-TEM appear as few-nanometers wide regions of continuous registry shift and often form dense networks in bilayer graphene.

**Grain boundary energies and out-of-plane deformations.** Formation energies play crucial role in determining the atomic structure of GBs at conditions close to thermodynamic equilibrium. This has been investigated theoretically using density functional theory[36] and empirical force fields[37,58,59]. Figure 2a shows the computed GB energies $\gamma$ for a number of symmetric periodic configurations characterized by different values of misorientation angle $\theta$[36]. Two scenarios can be considered here. First, GBs are constrained to assume flat morphology when strong adhesion of graphene to a substrate takes place. In this case, the energetics of these defects (filled symbols in Fig. 2a) can be described by the Read-Shockley equation as for bulk materials (solid line in Fig. 2a)[21,38]. The definition of misorientation angle $\theta$ given in Box 1 results in two small-angle regimes for which the distance $d$ between neighboring dislocations forming the GB is larger than the length of their Burgers vectors **b**. These regimes imply that $\gamma$ decreases as $d$ increases for $\theta \to 0°$ and $\theta \to 60°$, respectively. For intermediate values of $\theta$ the distance between neighboring dislocations is comparable to their Burgers vectors (large-angle GBs). Importantly, this regime is characterized by a minimum in $\gamma(\theta)$ (Fig. 2a). The low formation energies of large-angle GBs are explained by efficient mutual cancellation of in-plane elastic strain fields induced by closely packed dislocations. In particular, the two regular GB configurations shown in Box 1 have especially low formation energies of 0.34 eV/Å and 0.28 eV/Å, respectively, according to the results of first-principles calculations[36].

The situation of freely suspended graphene is remarkably different. Unlike bulk materials, the atoms of 2D graphene sheets are allowed to displace in the third, out-of-plane dimension. The possibility of out-of-plane displacement has profound effects on the energetics of topological defects in suspended graphene or graphene weakly bound to substrates. In particular, the out-of-plane corrugations effectively "screen" the in-plane elastic fields produced by topological defects thus dramatically reducing their formation energies[23,39]. While large-angle GBs in suspended graphene are flat, the stable configurations of small-angle defects are strongly corrugated (open symbols in Fig. 2a). Moreover, the out-of-plane displacements lead to finite magnitudes of otherwise diverging formation energies of isolated dislocations. First-principles and empirical force-field calculations predict formation energies of 7.5 eV[36] and 6.2 eV[60], respectively, for an isolated **b** = (1,0) dislocation. Remarkably, these values are comparable to formation energies of simple point defects in graphene, e.g. the Stone-Wales defect (4.8 eV) and single-atom vacancy (7.6 eV)[61]. The corrugation profile produced by a **b** = (1,0) dislocation appears as a prolate hillock (Fig. 2b), in agreement with the results of an STM study of dislocations in epitaxial graphene on Ir(111) substrate[62].

Out-of-plane deformations induced by the presence of topological defects in graphene have been investigated using electron microscopy techniques[63,64]. In TEM, the corrugation fields are observed indirectly as apparent in-plane compressive strain due to the tilting effect of graphene sheet. An example from Ref. [63] considers the case of a pair of dislocations separated by *ca*. 2 nm apart (Fig. 2c). The filtered image reveals the presence of an extended region of compressive strain connecting the two dislocations (Fig. 2d). Atomistic simulations assuming a perfectly flat graphene layer show only the presence of two localized in-plane stress dipoles (Fig. 2e), while allowing for out-of-plane relaxation reproduces the experimentally observed region of apparent compression (Fig. 2f). A 3D view of the out-of-plane deformation profile produced by a pair of dislocations is shown in Figure 2g.

Out-of-plane corrugation can also act as an efficient mechanism for relieving the misfit strain at asymmetric GBs in graphene. In this case, compressive strain was predicted to result in periodic ripples along the GB defects[40,65]. Such periodic ripples have recently been observed in an STM study of GBs on the surface of graphite[66].

**Transformations of topological defects.** Understanding the transformation pathways of topological defects is important for describing its plastic deformation. The motion of individual dislocations in graphene has been observed using HR-TEM[63,67]. In accord with early theory predictions, the two basic steps of dislocation motion – glide and climb – are realized by means of a single C–C bond rotation (the Stone-Wales transformation)[68,69] and removal of two carbon atoms[69,70], respectively. The energy barriers associated with these processes are sufficiently high to render them unlikely under equilibrium conditions. For instance, the energy barrier of a bond rotation step was predicted to lie in the 5–10 eV range[61]. Even higher energy barriers are expected for the sputtering of carbon atoms[71,72]. However, under TEM conditions, irradiation by high-energy electrons at accelerating voltages close to the displacement threshold (80 kV in Refs. 63,67) promotes the above-mentioned elementary processes of dislocation motion. In particular, both dislocation climb (Fig. 2h) and glide (Fig. 2i) have been observed. A complex glide process with an intermediate configuration involving an aggregate of three dislocations has also been evidenced (Fig. 2j).

Transformation of large-angle ($\theta \approx 30°$) GBs described as nearly continuous chains of pentagon-heptagon pairs was also investigated using TEM[52]. According to a simple thermodynamic argument, one expects a GB line to evolve only in the presence of significant boundary curvature. Indeed, nearly straight GBs showed fluctuating transformations without any time-averaged translation of the boundary line. In contrast, closed GB loops were shown to shrink under the electron irradiation leading to complete elimination of small graphene grains fully enclosed within another single-crystalline domain.

## CVD growth of polycrystalline graphene

While there are numerous ways to produce graphene, chemical vapor deposition (CVD) on polycrystalline Cu foils[35] has now become the most widely used method to synthesize high-quality, large-size monolayer graphene films due to its simplicity, low cost, and scalability. This technique produces the largest, over-meter-scale to date[73], graphene sheets which can be easily transferred to other substrates for diverse applications. The vast majority of experimental studies on GBs in graphene have been performed on such CVD grown samples. In such a CVD growth, thermal decomposition of hydrocarbon gas (most commonly $CH_4$, mixed with Ar and $H_2$) at high temperature provides the source of carbon atoms that will ultimately assemble into graphene on the surface of Cu substrate. Details of this process are subject to much research and are believed to involve multiple steps and intermediates[74,75]. Single-crystalline graphene grains nucleate around multiple spots (the nucleation centers) on the substrate, grow up in size, and as the growth proceeds, eventually merge to form a continuous polycrystalline graphene. Its properties will be determined by the constituent grains (their size, shape, edge orientation and other properties) and how they are merged or stitched together (i.e. the structure of GBs).

By stopping the growth before all the grains merge into a continuous polycrystalline film, single crystal grains (as well as isolated GBs between two grains) can be obtained,[35,76,77] allowing studying the formation and properties of these building blocks of polycrystalline graphene. It is noted that the polycrystallinity of the Cu foil is not a limiting factor for single-crystalline graphene growth, as a graphene grain can grow across GBs in Cu (Fig. 3a,b). This indicates weak interaction between graphene and Cu surfaces with no clear epitaxial relationship[76,77]. On the other hand, such interactions still exist and Cu crystal orientation can still have some influence on the growth of graphene overlayer[78,79,80,81,82,83]. Imperfections (defects, GBs and surface steps) and impurities in the Cu substrate can provide the nucleation centers for the growth[79,84]. Recently, it was discovered that the presence of oxygen on the Cu surface can substantially decrease the graphene nucleation density by passivating Cu surface active sites (Fig. 3c)[77]. Reducing the density of nucleation centers is the key to grow large single crystals of graphene[77,85,86,87]. Nucleation can also be artificially started using growth seeds[76,88].

Changing various growth parameters can control both the size and shape of graphene grains. For instance, grain size can be tuned by varying the growth rate[44]. Earlier studies noted that different CVD growth pressures can lead to different grain shapes, with the two most common ones being flower-shaped grains often obtained in low pressure (LP) CVD[35,76] and hexagonal shaped grains in atmospheric pressure (AP) CVD (Fig. 3a,b)[76]. The flower-like (dendritic) shape[35], with irregular and multifractal-like edges[35,89,90], indicates a diffusion-limited growth mechanism. The more regular hexagonal grains[76], whose edges are shown to be predominantly oriented along the zigzag directions of graphene lattice[76,91], represents an edge-attachment limited growth[77]. It was also realized that hydrogen plays important role by serving as an activator of the surface bound carbon needed in graphene growth as well as an etching reagent that controls the size and morphology of the graphene grains[92]. The shape and size of the grains can be continuously tuned by hydrogen partial pressure (Fig. 3d,e)[90,92]. Oxygen also accelerates graphene grain growth and shifts the growth kinetics from edge-attachment-limited (hexagonal shaped grains) to diffusion-limited (dendritic shaped grains) by reducing the activation barrier of the rate-limiting step[77]. The shape of grain is also affected by growth temperature[93]. In a growth mechanism model developed in Ref. 77, the shape is controlled by the balance between characteristic time of carbon attachment and carbon flux, with the longer attachment time favoring hexagonal shapes. Understanding the reactivity and kinetics of graphene edges is critical for understanding the growth mechanism[94,95].

## Optical Imaging and Characterizations

It is generally difficult for conventional optical imaging to directly visualize GBs in graphene, which typically have very narrow widths (far smaller than optical resolution) and do not have sufficient optical contrast with the surround graphene grains. We note that most GBs are also quite flat and can be difficult to see even with scanning electron microscopy (SEM) and atomic force microscopy (AFM)[76]. However, it is possible to utilize auxiliary agents that have characteristic interactions with graphene grains or GBs to facilitate optical imaging. For example, Kim *et al.* developed a simple method to visualize graphene grains and GBs by imaging the birefringence of a graphene surface covered with nematic liquid crystals, taking advantage of the orientation of the liquid crystals with underlying graphene lattice (thus mapping the graphene grains and GBs to those of the corresponding liquid crystal domains)[96]. Using optical microscopy, one can also enhance the contrast of grains and make GBs more visible by taking advantage of the more effective oxidization of substrate (Cu) in the region exposed or under GBs (which can be further functionalized to facilitate oxidization)[97,98]. Similar techniques can also be used to facilitate AFM and other types of GB imaging[99].

Beyond standard optical imaging, GBs can be easily visualized using spectroscopic Raman imaging of the defect-activated "D" mode (~1350 cm$^{-1}$)[76,97]. This also demonstrates strong inter-valley scattering of charge carriers at GBs (as such scattering is a part of the "D" peak Raman process) and corroborates the electronic transport measurements (*e. g.* weak localization) of GBs[76]. It is also possible to visualize and investigate electronic properties of GBs using an infrared nano-imaging technique as demonstrated in Ref. 100. They utilize surface plasmons that are reflected and scattered by GBs, thus causing plasmon interference patterns that are recorded to map out the GBs. Further quantitative analysis of such images reveals that GBs form electronic barriers (with effective width ~10−20nm, on the order of the Fermi wavelength in graphene and dependent on electronic screening) that obstruct both electrical transport and plasmon propagation[100]. The Raman and plasmonic based imaging do not require any auxiliary agents to treat the graphene.

**Transport phenomena**

Electronic transport properties of polycrystalline graphene are certainly among the most important considering potential applications of large-area graphene in electronics. Many experiments have shown, consistent with intuition, that GBs can impede electronic transport, thus degrading the conductive properties of polycrystalline graphene compared to single-crystalline graphene[76,101,102,103,104]. For example, mobility measurements of polycrystalline graphene samples showed larger grain sizes generally lead to higher electronic mobilities, despite substantial fluctuations in the values from sample to sample[101], while mobilities (either from Hall[76, 105] or field effect transistor[102] measurements) of single crystalline CVD graphene without GBs can exceed 10,000 cm$^2$/Vs and be comparable with exfoliated graphene[106]. Transport measurements on isolated individual GBs have shown that GBs generally result in enhanced electrical resistance (Fig. 4a,b), although the increase of resistance can vary across different GBs[76,105]. One the other hand, a few other experiments[18,44,107] observed no significant effects on the conductive properties due to GBs or variation of grain sizes, and some GBs that may even enhance the conduction (for example, GBs involving "overlapping" graphene layers[44], Fig. 4c). These different experimental results can be reconciled by considering that polycrystalline graphene samples may contain many types of GBs with different effects on electronic transport. Furthermore, when interpreting experimental results, one should keep in mind that typical graphene samples may contain many other types of disorder (such as point defects, contaminants, and nearby impurities) that can scatter electrons and lower the mobility, thus masking the effect of GBs.

Further transport measurements addressing dependence on temperature[102] and/or magnetic field[76,105] generated important insights into conduction and carrier-scattering mechanisms in polycrystalline graphene and helped understanding how GBs affect transport. For example, magnetotransport measurements across isolated GBs in comparison with those performed on single-crystalline grains have shown that GBs give rise to prominent weak-localization (WL, Fig. 4d) effects[76,105]. This indicates that GBs induce strong inter-valley scattering of carriers, consistent with the expected lattice disorder associated with GBs and with the observed strong Raman "D" peak (whose activation also requires inter-valley scattering[108,109]) for GBs[76]. Interestingly, it is also shown that the "half-integer" quantum Hall effect (QHE), which is a hall-mark of Dirac fermion transport in graphene, is maintained for the transport across GBs[76,105]. Well-developed QHE have been observed in polycrystalline samples as large as 1 cm in size[110].

Various scanning probe techniques have been employed as powerful tools to investigate the *local* electronic properties of GBs, down to the atomic scale, providing detailed insight into the microscopic mechanism of how GBs affect electronic transport. For example, both STM (including associated scanning tunneling spectroscopy (STS) that probes *dI/dV* conductance and local electronic density of states)[103] and conductive atomic force microscopy (c-AFM)[104] measurements have demonstrated suppressed conductivity of GBs. Multiple-probe STM measurements have found that the resistance of a GB changes with the width of the disordered transition region between adjacent grains[111]. Furthermore, STM revealed that GBs give rise to standing wave patterns[112] (Fig. 4e) propagating along the zigzag direction with a decay length of ~1 nm. This observation is indicative of backscattering and intervalley scattering processes (which were also observed at armchair graphene edges in graphene[91]), thus corroborating the Raman and weak-localization measurements[76,105]. In addition, STM/STS measurements have found that GBs tend to be more *n*-type doped[103,112] compared to the surrounding "bulk" graphene which is often found to be *p*-type doped due to surface adsorbates and contaminants. This leads to the formation of *p-n* junctions with sharp interfaces on the nanometer scale[103].

On the theory side, remarkable predictions have been reported for the periodic models of GBs investigated using the Landauer-Büttiker formalism[113]. Yazyev and Louie have shown that all periodic GBs can be divided into two distinct classes[65]. GB configurations belonging to the first class typically show very high probabilities of the transmission of low-energy charge carriers across the GB. Members of the second class, however, are characterized by significant transport gaps resulting in complete reflection of charge carriers in rather broad energy ranges (up to ~1 eV) around the charge-neutrality point (see Fig. 4f for schematic illustration of the effect). This striking transport behavior can be exploited in nanometer-scale electronic devices based on engineered GBs in graphene. Gunlycke and White have shown on the example of the structure described in Ref. 45 that transmission of charge carriers across periodic line defects at oblique angles can lead to their strong valley polarization (Fig. 4g)[114]. This prediction suggests that engineered line defects can be used as components of future valleytronic devices based on graphene[115,116,117]. The Landauer-Büttiker formalism has also been used for investigating the effect of strain[118] and chemical functionalization[119,120] on the transport properties of GBs in graphene. Electronic transport across disordered grain boundaries has been studied with the help of wavepacket evolution techniques either locally[121] or employing realistic large-scale models of polycrystalline graphene[122]. The latter work revealed a simple linear scaling of the mean free path and conductivity with the average size of single-crystalline domains. Several groups have proposed transport models based on the potential variation in the GB region. In particular, the effect of GBs on transport mobility was qualitatively explained using a potential barrier model[102]. A quantitative model of boundary resistance developed in Ref. 111 suggested the increased electron Fermi wave vector within the boundary region, possibly due to boundary-induced charge-density variation. Finally, the topological nature of intrinsic defects of polycrystalline graphene bears important implications for its electronic transport properties[123]. This effect of dislocations can be accounted by means of a gauge field[124,125] which gives rise to localized states at the Dirac point[126]. These localized states, in turn, result in resonant backscattering of low-energy charge carriers with the effect being especially strong for well-separated (i.e. belonging to small-angle GBs) dislocations[123].

## Mechanical properties

In bulk materials, the presence of defects usually leads to significant reductions of the tensile strength[127]. Knowing how dislocations and GBs affects mechanical properties of graphene is particularly important considering the facts that (i) single-crystalline graphene is the strongest known material[128] and that (ii) in low-dimensional materials the effect of disorder is expected to be amplified. In their recent systematic study, Lee *et al*. addressed mechanical properties of CVD-grown polycrystalline graphene[129]. Their technique relies on nanoindentation measurements of suspended graphene samples in an AFM setup as illustrated in Figure 5a. Elastic stiffness of polycrystalline graphene samples characterized for different grain sizes (1–5 $\mu$m and 50–200 $\mu$m) was found to be statistically identical to that of single-crystalline graphene. Figure 5b reproduces the histogram of the ensemble of AFM nanoindentation measurements performed on large-grain (50–200 $\mu$m) samples. The obtained elastic modulus of 339 ± 17 N/m corresponds to a 3D Young's modulus of ~1 TPa. The measured fracture loads of large-grain polycrystalline samples were also found statistically identical to those of pristine single-crystalline graphene (Fig. 5c). However, sizable reduction of the mean value and broader distribution of fracture loads were observed for small-grain (1–5 $\mu$m) samples. This implies that the strength of polycrystalline graphene is affected by randomly occurring structural defects. Indeed, nanoindentation measurements performed directly on GB showed 20%-40% smaller fracture loads compared to those performed in the middle of single-crystalline domains (Fig. 5d). Investigations of graphene membranes after indentation shows that cracks propagate not only along GBs, but also inside grains (Fig. 5e). These results suggest that the elastic stiffness and strength of polycrystalline graphene with well-stitched GBs are close to those of pristine graphene. Importantly, the study also brings to one's attention the fact that post-growth processing techniques used in the previous studies[130] significantly degraded the strength of graphene.

On the theory side, a number of computational studies of mechanical properties of polycrystalline graphene have been reported. Simulations performed on single GB models reveal that large-angle defect configurations (like the ones shown in the right column of Box 1) are as strong as pristine graphene, while small-angle defects characterized by larger inter-dislocation distances are significantly weaker[131,132]. We note that a recent nanoindentation study by Rasool *et al.* shows a clear reduction of the fracture force when small-angle GBs are probed[133]. The computational studies also provide an insight into the atomistic picture of mechanical failure of polycrystalline graphene. In particular, several recent works reported simulations performed on models containing realistic GB networks[134,135,136]. It was found that triple junctions of GBs serve as nucleation center for cracks. Propagation of cracks both along GBs and within single-crystalline grains was observed (Fig. 5f)[134,135]. Importantly, the failure of graphene can be considered as brittle since dislocations are completely immobile at normal conditions[135,136].

In additional to mechanical properties, it is also interesting to consider whether and how GBs may affect or even be used to engineer phonon (thermal) transport[137,138]. This remains a largely unexplored topic so far, calling for much further studies, especially experiments. The possible different ways GBs may affect thermal transport compared to electronic transport could promise interesting functionalities, e. g. for thermoelectric devices.

**Topological defects in other 2D materials**

From structural point of view, h-BN and TMDCs are closely related to graphene (Fig. 6a). Monolayer *h*-BN has the same honeycomb crystal structure as graphene, with a similar lattice constant (a = 2.50 Å) and two sublattices populated by B and N atoms, respectively. Such a polarity of crystalline lattice leads to profound consequences when atomic structure of topological defects is considered. In particular, odd-membered rings present in the topological defects break the alternating order of the two atomic species resulting in homoelemental bonding[139]. The presence of covalent bonds between the likely charged atoms increases formation energies of defects and introduces an extra degree of freedom in defining their structures. For example, the **b** = (1,0) dislocations now come in two different flavors featuring either a B–B bond or an N–N bond (Fig. 6b). The two possible structures of dislocations cores are characterized by local deviation from the nominal stoichiometry (B-rich and N-rich) and by local charges (positively and negatively charged dislocations, respectively). Alternatively, dislocation cores involving only even-membered rings would allow avoiding the formation energy penalty associated with homoelemental bonding. An example of such neutral dislocation core composed of an edge-sharing pair of 4- and 8-membered rings is shown in Figure 6b. However, this dislocation is characterized by a larger Burgers vector **b** = (1,1) implying a significantly higher contribution to the formation energy due to elastic field. Available experimental evidence speaks in favor of defect structures containing homoelemental bonds. Atomic resolution TEM image of a GB in monolayer *h*-BN shows three dislocations with cores composed of pentagon-heptagon pairs, two of which are B-rich and one is N-rich (Fig. 6c)[140].

The crystal structure of monolayer TMDCs is composed of triangular lattice of metal atoms sandwiched between two triangularly packed planes of chalcogen atoms. In the most stable 2H phase, chalcogen atoms are stacked along the direction orthogonal to the monolayer. The crystal structure is effectively a 2D honeycomb lattice with the two sublattices populated by metal atoms and the pairs of chalcogen atoms (Fig. 6a). In this structure, metal atoms are 6-fold coordinated while chalcogen atoms are 3-fold coordinated. Compared to the case of *h*-BN, however, these 2D materials are expected to display even greater structural variety of topological defects because of (i) more easily changed coordination number of the constituent atoms, and (ii) the 3D character of their coordination spheres[141,142]. The binary chemical composition of *h*-BN and TMDCs also implies that topological defects in these materials realize different stoichiometries which makes their formation energies dependent on the chemical potential. Potentially, the chemical degree of freedom presents another opportunity for engineering topological defects in 2D materials.

Several experimental studies of polycrystalline monolayer $MoS_2$ have been published recently[143,144,145]. The samples have been produced by CVD growth starting from solid $MoO_3$ and S precursors. This growth procedure results in triangular islands (single crystalline domains) of monolayer $MoS_2$ with well-defined edges oriented either along Mo-rich or along S-rich zigzag directions (Fig. 6d). Growing islands inevitably form polycrystalline aggregates as exemplified in Fig. 6e. The images reported by van der Zande *et al.* reveal the highly-faceted morphology of grain boundaries with their preferred crystallographic orientation inclined by *ca.* 20° with respect to the zigzag direction[143]. At the atomic scale, the interface is composed of a regular arrangement of 4- and 8-membered rings as shown in Fig. 6f. It was shown that such GBs have strong effects on optical properties and electrical conductivity consistent with the theoretical predictions of mid-gap states induced by such defects. Interestingly, somewhat different defect structures were reported by other authors in the samples of monolayer MoS$_2$ produced by a very similar CVD process[144,145]. The 60° GBs were shown to be composed predominantly of 4-membered rings arranged in two different patterns (Fig. 6g,h)[144]. Moreover, the observed small-angle GBs allowed to identify individual **b** = (1,0) dislocations. Their great structural diversity summarized in Figure 6j shows dislocation cores featuring 4-membered rings, Mo–Mo and S–S bonds as well as undercoordinated S atoms.

Atomically thin silica comprises another interesting example of hexagonal 2D lattice formed by corner-sharing SiO$_4$ tetrahedra. TEM investigations of bilayer silica reveal the presence of topological and point defects similar to those observed in graphene[146,147]. Unlike graphene, however, 2D silica realizes an extra degree of structural freedom associated with the rotation of individual SiO$_4$ units. This results in additional relaxation effects leading to very low formation energies of certain defects such as the large-angle GBs[148]. This peculiarity is ultimately reflected in strong tendency of 2D silica to form vitreous (amorphous) phase, which realizes the limit of high concentration of topological defects[149,150].

## Perspectives

The overview of recent progress highlights priority directions of research in the field of polycrystalline 2D materials. While significant progress has been achieved in understanding the mechanism of CVD growth of graphene, much remains unclear regarding how graphene grains merge during the growth as their edges evolve and interact and what defines the details of the atomic-scale structure of GBs formed. Significant research efforts will be devoted to developing methods of post-growth modification of polycrystalline graphene aimed at achieving desirable properties. Optimizing large-scale growth processes for increasing the size of single-crystalline graphene will remain one of the major vectors of research. However, tailoring graphene, its electrical, thermal, thermoelectric, mechanical and chemical properties, by means of purposefully introducing and manipulating topological disorder is expected to become another important objective. Atomic-precision engineering of individual topological defects for using them as components of novel nanoscale devices sets the ultimate challenge for researchers. On the theory side, a lot remains to be done in terms of developing new tools for multiscale simulations of realistic models of polycrystalline 2D materials. Understanding the relations between the structural topological invariants of intrinsic defects in polycrystalline materials and their effects on various properties of these materials will continue stimulating research. Finally, as the family of 2D materials continues expanding, the issues of their polycrystalline nature and related topological defects specific for each of these novel materials will have to be addressed.

**Acknowledgements**

O.V.Y. acknowledges funding from Swiss NSF (grant No. PP00P2_133552) and ERC Grant "TopoMat" (No. 306504) and technical assistance of G. Autès and F. Gargiulo in preparing the manuscript. Y.P.C. acknowledges research support from NSF, NIST, DTRA, DHS and Purdue University and collaborations and helpful discussions with many group members and colleagues on topics related to this review, in particular H. Cao, T.-F. Chung, R. Colby, L. A. Jauregui, E. A. Stach, J. Tian and Q. Yu.

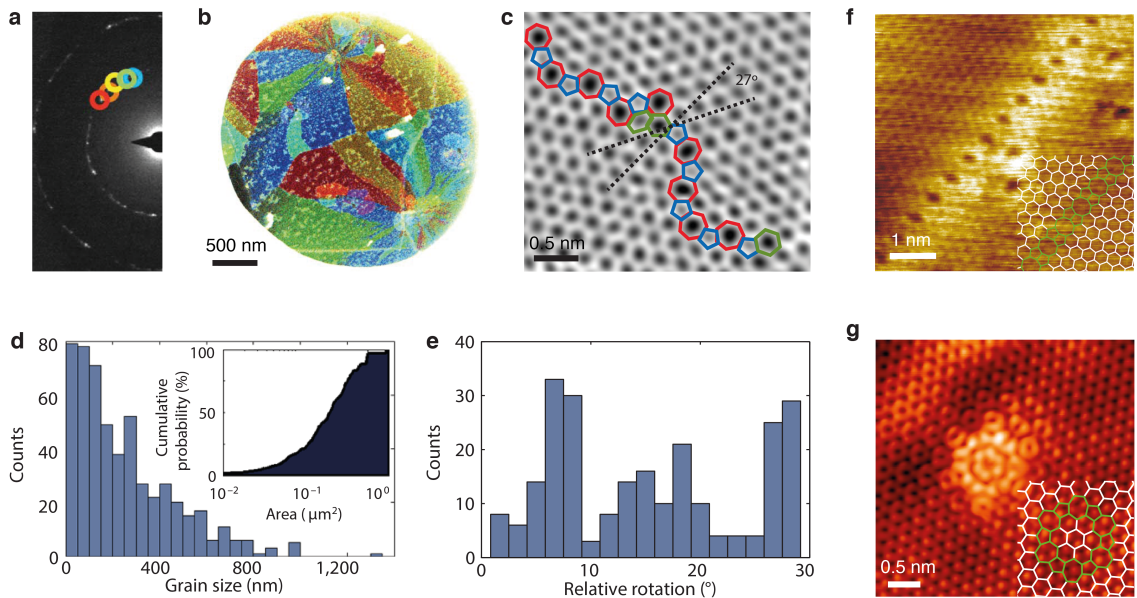

**Figure 1 | Experimental studies of polycrystalline graphene and extended defects. a,** Electron diffraction pattern from a sample of polycrystalline graphene showing numerous sets of 6-fold-symmetric diffraction spots rotated with respect to each other. **b,** False-colored DF-TEM image revealing individual single-crystalline graphene grains of varying shape, size and orientation. This image was constructed by aperturing the diffraction spots in **a** such that only the scattered electrons corresponding to one set of diffraction spots (color-coded circles in panel **a**) is used to construct the real-space image. **c**, Aberration-corrected annular DF-scanning TEM (STEM) image of a GB stitching two graphene grains with lattice orientations rotated by ~27° with respect each other. The dashed lines outline the lattice orientations of the two domains. The structural model highlighting heptagons (red) and pentagons (blue) is overlaid on the image. **d,e,** Distributions of grain sizes and relative orientations, respectively, in samples of polycrystalline graphene investigated in Ref. 18. **f,** STM image of a regular line defect in graphene grown on Ni(111) substrate[45]. Inset shows the structural model. **g**, STM image of the flower-shaped point defect in epitaxial graphene grown on SiC(0001)[48]. Inset shows the structural model. Figures reproduced with permission from: **a−e**, ref. 18, © 2011 NPG; **f**, ref. 45, © 2010 NPG; **g**, ref. 48, © 2011 APS.
32

**Box 1 | Relations between different types of topological defects in graphene.**

Polycrystalline materials are composed of single-crystalline domains with different lattice orientations. The changes of the lattice orientation are accommodated by the presence of topological defects. The structure of such defects is described by some topological invariant, a non-locally defined quantity conserved upon local structural transformations. There are three types of topological defects relevant to 2D materials − disclinations, dislocations and grain boundaries (GBs) – related to each other by hierarchical relations[23,36,39]. Importantly, in graphene these defects can be constructed without perturbing the native three-fold coordination sphere of $sp^2$ carbon atoms[36].

**Disclinations** (**a**) are the elementary topological defects obtained by adding a semi-infinite wedge of material to, or removing from, an ideal 2D crystalline lattice. For 60° wedges, the resulting cores of positive ($s = 60°$) and negative ($s = -60°$) disclinations are pentagons (red) and heptagons (blue), respectively, embedded into the honeycomb lattice of graphene. Wedge angle σ is the topological invariant of a disclination. The presence of isolated disclinations in graphene, however, is unlikely as it inevitably results in highly non-planar structures.

**Dislocations** (**b**) are the topological defects equivalent to pairs of complementary disclinations. The topological invariant of a dislocation is the Burgers vector **b** which is a proper translation vector of the crystalline lattice. A dislocation effectively embeds a semi-infinite strip of material of width **b** into a 2D lattice[36]. An edge-sharing heptagon-pentagon is a dislocation in graphene with the smallest possible Burgers vector equal to one lattice constant (**b** = (1,0)). Larger distances between disclinations result in longer Burgers vectors as illustrated by the **b** = (1,1) dislocation.

**Grain boundaries** (**c**) in 2D materials are equivalent to 1D chains of aligned dislocations[38]. These topological defects are the ultimate interfaces between single-crystalline grains in polycrystalline materials. The topological invariant of a GB in 2D is the misorientation angle $\theta = \theta_L + \theta_R$ ($0° < \theta < 60°$), which is related to the density of dislocations and their Burgers vectors **b** via the so-called Frank's equations[21]. Large dislocation density (or, equivalently, small distance between the neighboring dislocations) corresponds to large misorientation angles. Two examples of particularly stable large-angle GBs ($\theta = 21.8°$ and $\theta = 32.3°$) in graphene are shown.

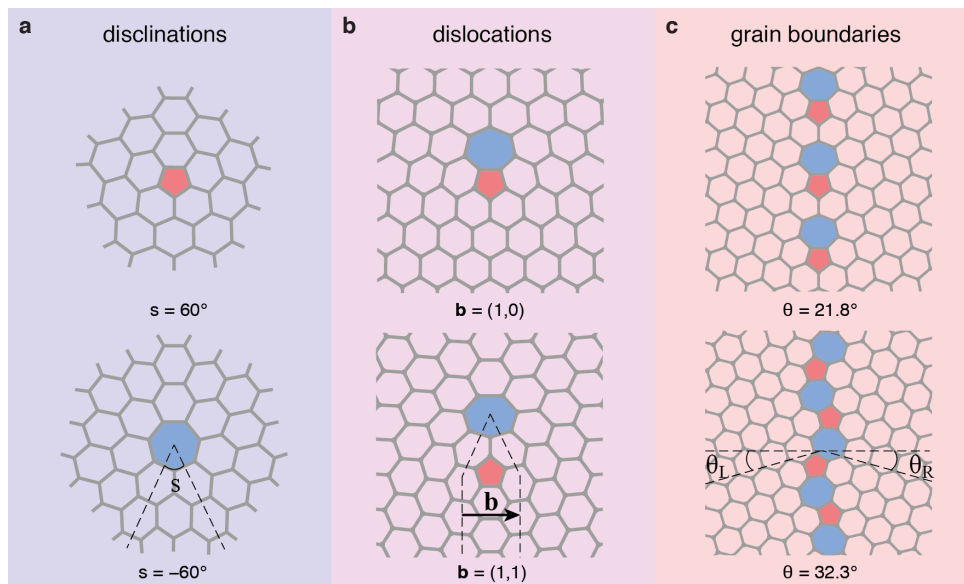

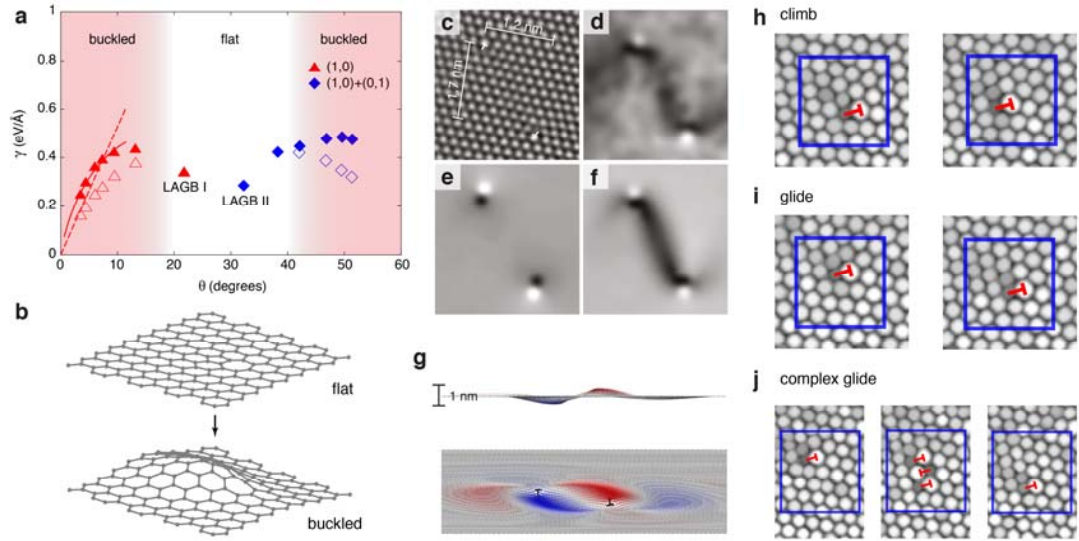

**Figure 2 | Out-of-plane deformations and transformations of topological defects. a,** Grain-boundary energies $\gamma$ plotted as a function of misorientation angle $\theta$ for symmetric defect configurations[36]. The color of symbols reflects the Burgers vectors of constituent dislocations. Solid and open symbols correspond to flat and buckled configurations, respectively. Solid and dashed lines correspond to the fits assuming the Read-Shockley equation and the finite formation energy (7.5 eV) of dislocations. Shaded areas indicate the ranges of misorientation angle in which the buckled configurations are energetically preferred over the flat ones. **b,** Transition to an out-of-plane corrugated state of graphene sheet produced by the presence of a **b** = (0,1) dislocation. **c**, HR-TEM image of a pair of **b** = (0,1) dislocations in graphene separated by 1.2 nm glide distance and 1.7 nm climb distance. **d**, Filtered image reveals the apparent in-plane compression (dark) and extension (bright). Panels **e** and **f** show simulated filtered images corresponding to flat and buckled configurations, respectively. **g**, Lowest energy configuration of the corrugation produced by a pair of dislocations in relative arrangement similar to the one shown in panel **c**. Out-of-plane displacements of carbon atoms are color-coded. **h**,**i**, Maximum filtered HR-TEM images reveal the dislocation climb and glide processes, respectively. **j**, An observation of complex glide process which starts by a bond rotation event next to the dislocation core and involves an intermediate aggregate of 3 dislocations. Figures **c**−**j** reproduced with permission from ref. 63, © 2013 NPG.

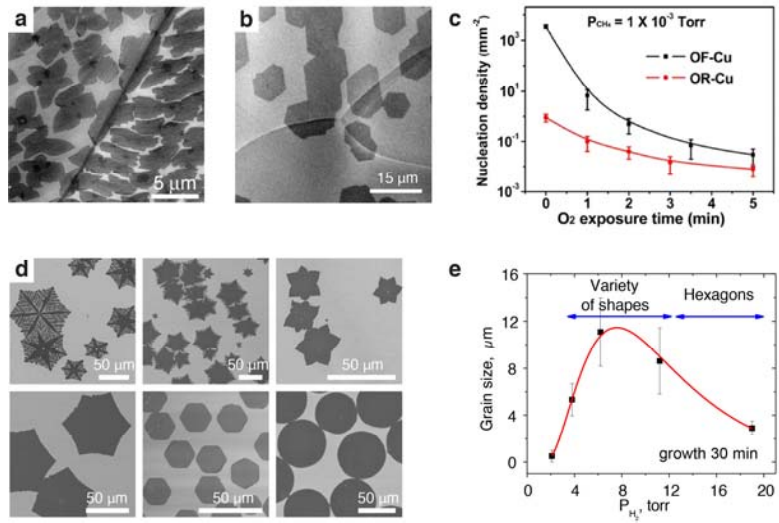

**Figure 3 | Growth of graphene grains and polycrystalline graphene by chemical vapor deposition (CVD). a,** Flower-shaped and **b,** hexagonal-shaped graphene grains nucleated from different locations on Cu foil in LP-CVD (**a**) and AP-CVD (**b**) growth, stopped before the grains merge into a continuous sheet of polycrystalline graphene. It is seen that graphene grains can grow continuously across Cu GBs on polycrystalline Cu foils. **c,** Nucleation density of graphene grains measured as a function of $O_2$ exposure time on oxygen-free (OF-Cu) and oxygen-rich (OR-Cu) Cu foils demonstrating the effect of $O_2$ to reduce nucleation density and increase the grain size[77]. **d,** Evolution of the shape of graphene grains (from snow-flake, flower-like, to hexagonal, and even circular) with varying growth conditions (mainly increasing $H_2$:Ar ratio, with some adjustments in $CH_4$ and growth time) in an AP-CVD growth on melted Cu foil[90]. **e,** Effect of $H_2$ partial pressure during CVD growth on the size and shape of graphene grains, where flower-like or irregular shapes are obtained at low and hexagonal shape at high $H_2$ partial pressure[92]. Similar phenomena are observed in both AP-CVD and LP-CVD. Images in panels **a**, **b** and **d** are obtained with scanning electron microscopy (SEM). Figures reproduced with permission from: **a**, ref. 35, © 2009 AAAS; **b**, ref. 76, © 2011 NPG; **c**, ref. 77, © 2013 AAAS; **d**, ref. 90, © 2013 NPG; **e**, ref. 92, 2011 © ACS;

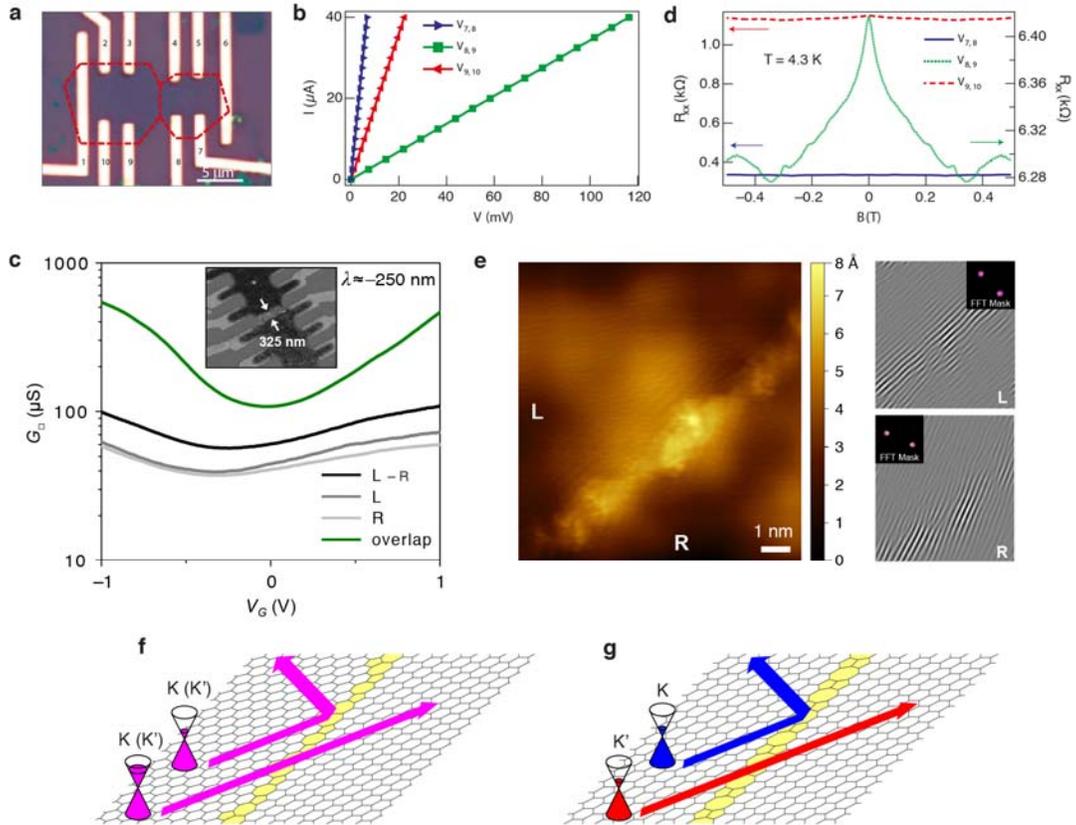

**Figure 4 | Electronic transport across grain boundaries in graphene. a,** Image of device from ref. 76 with red dashed line indicating the contour of two coalesced graphene grains with a GB in between. **b,** Room-temperature current-voltage ($I$–$V$) curves measured across an isolated GB (green) and those within the graphene grain on each side of the GB (red and blue). This measurement demonstrates a significantly reduced conductance (slope of the $I$–$V$ curve) of cross-GB transport compared to intra-grain transport[76]. **c,** Gate voltage ($V_G$) dependent sheet conductance ($G_{\square}$) measured across an "overlapped" GB (labeled "L-R") and within the grain on each side of the GB ("L" and "R"), showing that such a GB actually enhances the conductance (or effective "shortens" the transport path length, in this case by ~250nm)[44]. The conductance contributed by the GB is extracted (green). Inset shows device image with the width of "overlap" region indicated (325 nm). **d,** Low-temperature magnetoresistance measured across the GB and within the grains for the device shown in panel **a**. The measurement reveals weak localization behavior indicative of inter-valley scattering caused by the GB. **e,** STM image around a GB, revealing a linear structure (more clearly seen in the Fourier-filtered images in inset) that is associated with GB-induced intervalley scattering[112]. **f,** Schematic illustration of the transport gaps predicted for certain periodic GB defects in Ref. 65. While low-energy charge carriers are perfectly reflected, transmission across the GB defect is enabled at higher energies. **g,** Schematic illustration of the valley filtering effect predicted for the line defect in Ref. 114. At oblique incidence angles charge carriers transmitted across the line defect acquire strong valley polarization. The structures of defects in panels **f** and **g** are highlighted. Figures reproduced with permission from: **a,b,d**, ref. 76, © 2011 NPG; **c**, ref. 44, © 2012 AAAS; **e**, ref. 112, © 2012 ACS.

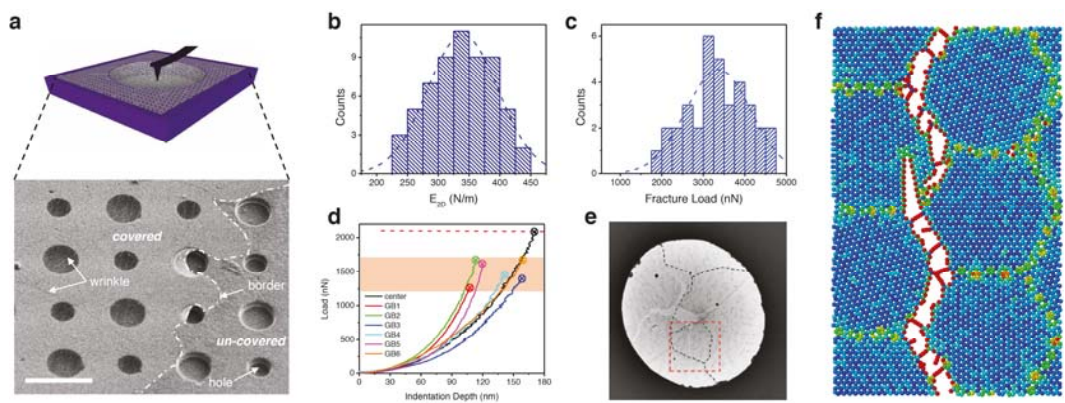

**Figure 5 | Mechanical properties of polycrystalline graphene. a**, Scheme of an AFM nanoindentation setup for probing mechanical properties of suspended graphene used in Ref. 129. **b**,**c**, The histograms of elastic stiffness (**b**) and fracture load (**c**) measured for polycrystalline graphene samples characterized by a large grain size (50–200 μm). The dashed lines represent fitted Gaussian distributions. **d**, Force-displacement curves for measurements performed on the GB and in the middle of a single-crystalline domain. **e**, Bright-field TEM image of cracks in suspended graphene after the nanoindentation experiment. The black dashed lines indicate location of GBs before the nanoindentation experiment. **f**, Snapshot of a molecular dynamics simulation of the fracture process performed on an atomistic model of polycrystalline graphene[135]. The simulations reveal fracture via propagation of both inter- and intragranular cracks. Figures reproduced with permission from: **a−e**, ref. 129, © 2013 AAAS; **f**, ref. 135, © 2012 ACS.

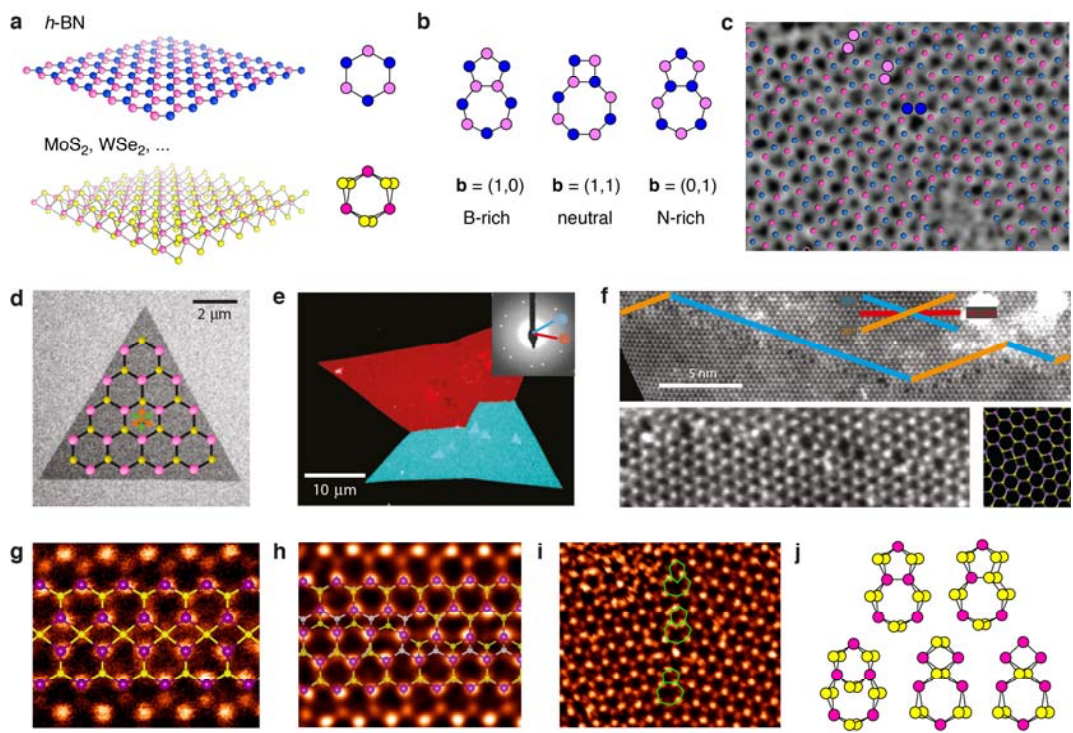

**Figure 6 | Grain boundaries in binary 2D materials. a**, Structural models of binary 2D materials – monolayer hexagonal boron nitride (*h*-BN) and transition metal dichalcogenides ($MoS_2$, $WSe_2$ and related materials). **b**, Proposed structures of dislocations in *h*-BN[139]. The **b** = (1, 0) dislocation cores composed of a pair of 5- and 7-membered rings are charged and include a homoelemental bond. The neutral **b** = (1, 1) dislocation core involves 4- and 8-membered rings. **c**, HR-TEM image of a GB in *h*-BN showing the presence of both types of charged dislocations[140]. The atoms belonging to homoelemental bonds at dislocation cores are indicated. **d**, Single crystal of monolayer $MoS_2$ with Mo-rich edges oriented along zigzag direction. **e**, DF-TEM image of a polycrystalline aggregate of $MoS_2$. The color-coding corresponds to the diffraction spots indicated in the inset. **f**, Atomically resolved image (annular DF (ADF)-STEM) of a GB in $MoS_2$ reveals its highly faceted structure and preferential crystallographic orientation[143]. The interface is composed of a regular arrangement of 4- and 8-membered rings, as shown in the magnified image and structural model. **g**,**h**, Highly regular structures of 60° GBs in $MoS_2$ observed in Ref. 144. **i**, High-resolution image (ADF-STEM) of an 18.5° GBs consisting of dislocations with 5- and 7-membered rings. **j**, Structural models of experimentally observed dislocation cores in monolayer $MoS_2$ reported in Refs. 144 and 145. Figures reproduced with permission from: **c**, ref. 140, © 2013 ACS; **d**−**f**, ref. 143, © 2013 NPG; **g**−**i**, ref. 144, © 2013 ACS.